\documentclass[
    ,final            
  ]
  {aipproc}

\layoutstyle{6x9}

\begin{document}

\title{Heavy Hyperon--Antihyperon Production}

\classification{13.75.Cs, 14.20.Jn, 25.43.+t}
\keywords      {Heavy Hyperon - Antihyperon, Systems out of 3$\bar s s$}

\author{W. Oelert}{
  address={IKP -- Research Centre J\"ulich, D--52425 J\"ulich, Germany}
}

\author{D. Grzonka}{
  address={IKP -- Research Centre J\"ulich, D--52425 J\"ulich, Germany}
}

\author{L. Jarczyk}{
  address={Institute of Physics, Jagellonian University, PL--30--059 Cracow, Poland}
}

\author{K. Kilian}{
  address={IKP -- Research Centre J\"ulich, D--52425 J\"ulich, Germany}
}

\author{P. Moskal}{
address={Institute of Physics, Jagellonian University, PL--30--059 Cracow, Poland}  
}
\author{P. Winter}{
  address={IKP -- Research Centre J\"ulich, D--52425 J\"ulich, Germany}
}
\begin{abstract}
 Based on the experience from the production of light antihyperon--hyperon 
($\bar \Lambda \Lambda,~ \bar \Sigma \Sigma$) pairs at LEAR (experiment
PS185) it is suggested to continue the investigations towards the
heavier antihyperon--hyperon pairs $\bar \Xi \Xi$ and 
$\bar \Omega \Omega$
in view of:
\begin{enumerate}
\item
the production dynamics of the heavier antihyperon--hyperon out of the $\bar p p$ 
annihilation\\[-0.2cm]
\item 
a comparison of the 
''(3 $s$) (3 $\bar s$)''-- quark system $\bar \Omega \Omega$ \\
to the
3 ($\bar s s$) = 3 $\phi$ meson production,\\
where both systems the $\bar \Omega \Omega$ and the 3 $\phi$
have similar masses \\
(3.345 and 3.057, respectively) and identical valence quark content. 
\end{enumerate}
A systematic study of the antihyperon--hyperon production with increasing
strangeness content is interesting for the following reasons:
The $\bar \Omega \Omega$ production is the creation of two
spin 3/2 objects out of the two spin 1/2 $\bar p p$ particles. Results of the PS185
experiments prove a clear dominance of the spin triplet $\bar s s$ dissociation.
In the $\Omega $($\bar \Omega$) the three $s$-quarks 
(three $\bar s$-quarks) are aligned to spin 3/2 each. If the three  $\bar s s$
pairs are now all in spin triplet configurations when created out of the gluonic 
interaction they should have spin parity quantum
number as $3^-$ as long as $\Omega $$\bar \Omega$ is created with relative $L = 0$
angular momentum. \\
The comparison of the $\Omega $$\bar \Omega$ baryon pair to
the $\phi \phi \phi $ three meson production (where the three $\bar s s$ quark
pairs
might not but can be produced without relative correlation) would provide
a unique determination of the intermediate matter state.\\
Measurements of excitation functions and polarization transfers should be used
to examine these gluon rich $\bar p p \to \bar \Omega \Omega$ and 
$\bar p p \to \phi \phi \phi$ reaction channels.\\
Such experiments should be performed at the PANDA detector at the FAIR facility of the GSI.
 
\end{abstract}

\maketitle
\section{Introduction}
Interest for studying the production of hyperon--antihyperon pairs following
antiproton--proton annihilations is based largely on the aim to understand the
nature of flavour production and its dynamics. There is a large amount of data from the
PS185~\cite{PS185} collaboration as reported during the LEAP-05 conference by
T.~Johansson~\cite{tord} and J.M.~Richard~\cite{JMR}. The goal of those PS185
experiments is to establish the definitive $\bar p p \to \bar Y Y$  data set in
the low--energy regime. From the theoretical analysis of the data, one hopes to gain
insight into the behaviour of hadron interactions at intermediate energies, i.e.
in an energy regime where perturbative QCD is inappropriate and both
quark--gluon and meson degrees of freedom are believed to be important.\\
Heavy hyperon--antihyperon production studies are relevant since i) rather little
is known about heavy flavour hyperons, even down to the final proof of their
existence; ii) different hyperon--antihyperon pairs are produced in distinct but
specific hadronic environments; iii) threshold production studies keep the
interpretation as simple as possible and iv) quark dynamic effects can be
observed. The special interest for the investigation of $\bar \Omega \Omega$
production will be discussed in this contribution.\\
\section{Results from the PS185 experiment} 
Results of the PS185--collaboration studies --~as far as relevant for the recent
discussion~-- are shown in figure~1 (left) and (right), demonstrating that the differential
cross sections for the hyperon--antihyperon pair production following the $\bar p
p$ annihilation feature the onset of higher partial waves already at very low
excess energies  and that the singlet fraction for the   $\bar p p \to \bar \Lambda \Lambda$ reaction
is largely consistent with zero, respectively.\\
\begin{figure} [h]
\hspace{0.0cm}
  \includegraphics[height=.24\textheight]{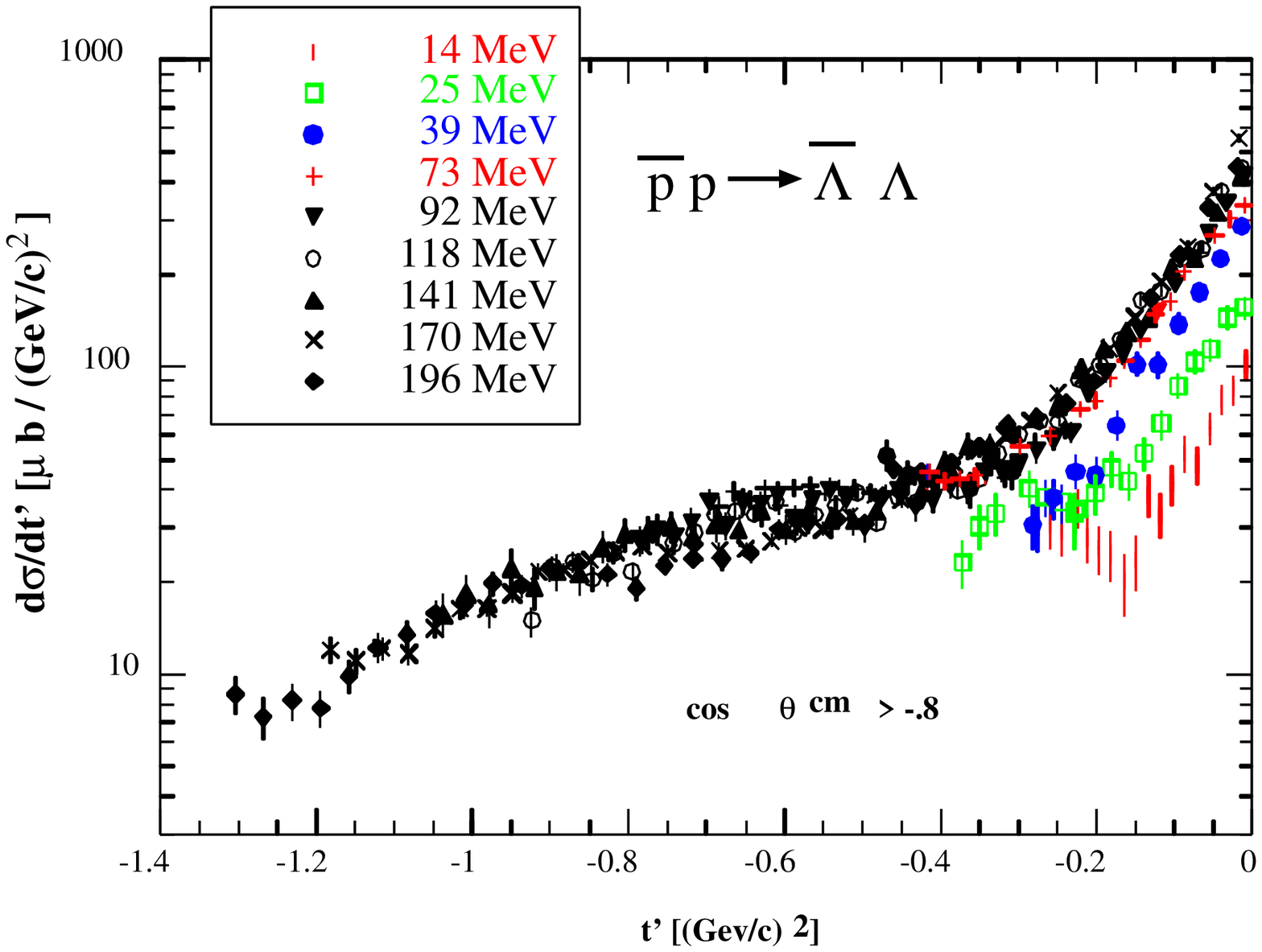}
\hspace{0.0cm}
\vspace{+7.0cm}
  \includegraphics[height=.24\textheight]{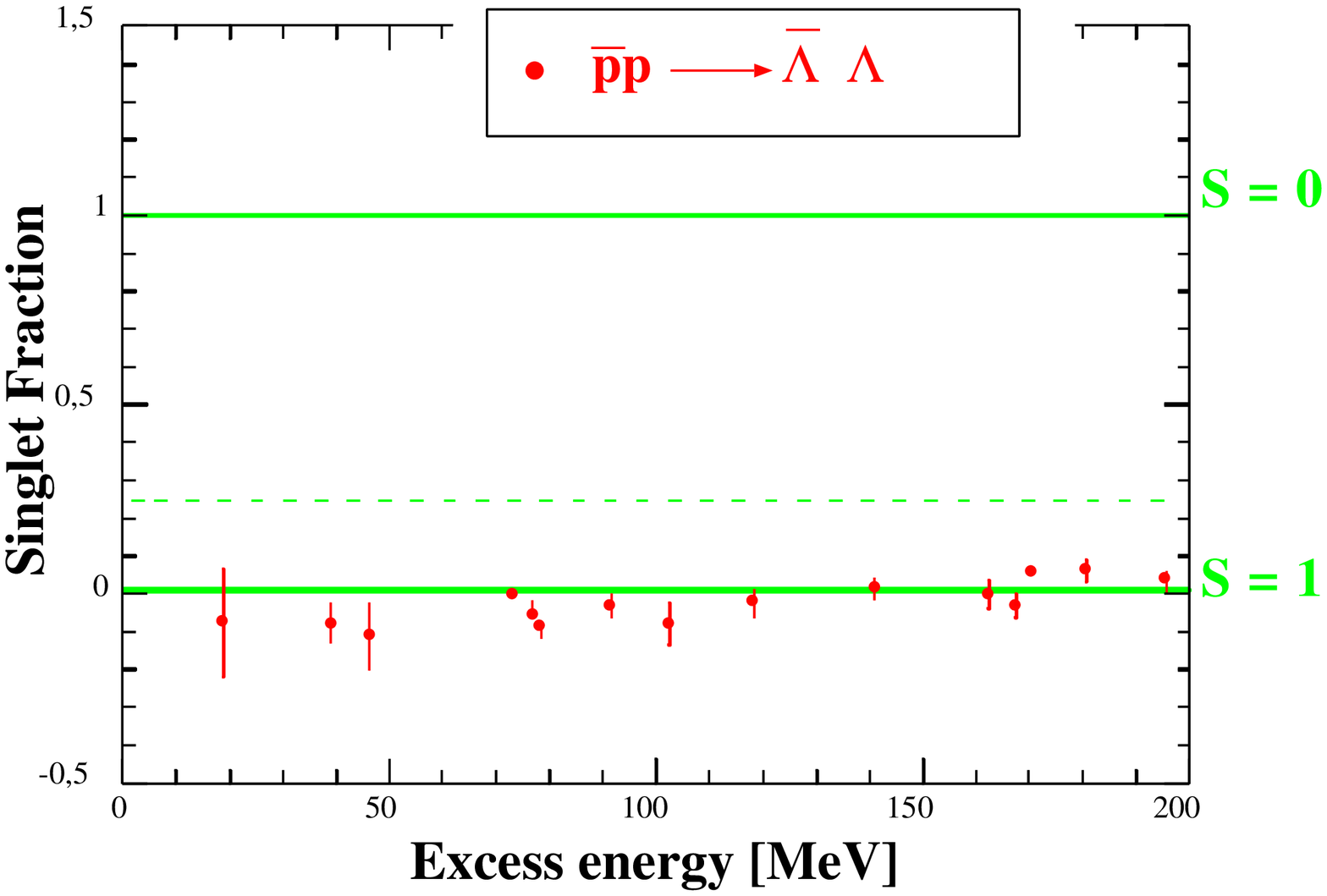}
  \caption{Left: PS185 results of differential cross sections at various energies, shown
  as a function of momentum transfer. $~$ 
  Right: PS185 results of the singlet fractions for the reaction:
  $\bar p p \to \bar \Lambda \Lambda$   
  at various energies, shown
  as a function of the excess energy}  
\end{figure}
\section{From light to heavy hyperons}
\begin{figure}[htp]
\hspace{1.00cm}
\includegraphics[width=1.11\textwidth]{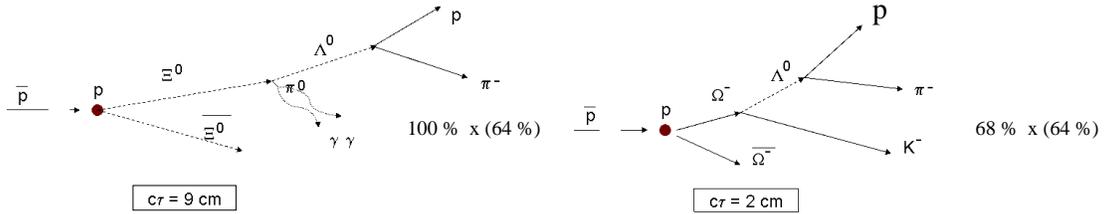}
\caption{Decay scheme of the two hyperons $\Xi^0$ and $\Omega^-$}
\label{fig:H_D}
\vspace{-0.50cm}
\end{figure}

Based on the experimental and theoretical experience from the production of light antihyperon--hyperon pairs at
LEAR it is suggested to measure the production of both heavy antihyperon--hyperon pairs
and the three $\phi$-meson system at the future facility FAIR at GSI. The foreseen
momentum range~\cite{FAIR_LOI} of the $\bar p$--beam at the HESR/FAIR allows to study hyperons with one, two and
three valence s-quarks or one valence c-quark up to the $\Xi_c$ with its mass of M
= 2.47 GeV/c$^2$ in the $\bar p p$ interaction, see table 1.\\
\begin{table}
\begin{tabular}{crrr}
\hline
  \tablehead{1}{r}{b}{Hyperon\\    }
  & \tablehead{1}{r}{b}{Mass\\~~~~~~(MeV/c$^2$)}
  & \tablehead{1}{r}{b}{$\sqrt s$\\~~~~~~(MeV)}
  & \tablehead{1}{r}{b}{Beam \\~~~~~~~~Momentum\\(GeV/c)}   \\
\hline
$\Lambda~~~$ &~~~~~~~~~~~~~~~~~~1115.57 $\pm$ 0.06 &~~~~~~~~2231.14 & 1.435    \\
$\Sigma^+~$ &1189.37 $\pm$ 0.06 &2378.74 &1.854       \\
$\Sigma^0~~$ &1192.55 $\pm$ 0.10 &2385.1~~ &1.871       \\
$\Sigma^-~$ &1197.50 $\pm$ 0.05 &2395.0~~ &1.900       \\
$\Xi^0~~$     &1314.80 $\pm$ 0.8~~ &2629.60 &2.582       \\
$\Xi^-~~$     &1321.34 $\pm$ 0.14 &2642.68 &2.621       \\
$\Omega^-~~$  &1672.43 $\pm$ 0.14 &3344.86 &4.936       \\
\\
$\Lambda_c^+~~$ &2285.2~~ $\pm$ 1.2~~ &4570.4~~ &10.150       \\
$\Sigma_c^{0}~~$ &2452.7~~ $\pm$ 1.3~~ &4905.4~~ &11.848       \\
$\Sigma_c^{++}$ &2453.0~~ $\pm$ 1.2~~ &4906.0~~ &11.851       \\
$\Sigma_c^{+}~$ &2453.2~~ $\pm$ 1.2~~ &4906.4~~ &11.853       \\
$\Xi_c^{+}~~$ &2466.5~~ $\pm$ 2.5~~ &4933.0~~ &11.993       \\
$\Xi_c^{0}~~$ &2473.1~~ $\pm$ 2.0~~ &4945.2~~ &12.063       \\
$\Omega_c^{0}~~$ &2740.~~~~ $\pm$ 2.0~~ &5480.~~~~ &15.1       \\[0.1cm]
\hline
\\[-0.5cm]
\end{tabular}
\caption{{\mbox{Relevant parameters for the production of hyperons in  $\bar p p$
annihilation}}}
\label{tab:a}
\end{table}
\\
The common special feature of the hyperons is their weak (flavour changing)
decay with $c\tau$ values of a few centimeters, as presented in table~2.
Typical decay schemes 
of the lighter hyperons might be represented in two categories. 
The charged hyperons (antihyperons) decay mainly~(see~figure~\ref{fig:H_D}) delayed into a 
charged meson and a $\Lambda$ ($\bar \Lambda$) which then decays to
the two charged particles $\pi^-$ and $p$ ($\pi^+$ and $\bar p$). 
The neutral hyperons (antihyperons) decay via $\Lambda + \pi^0$
($\bar \Lambda + \pi^0$) into the two charged particle system $\pi^-$ and $p$
($\pi^+$ and $\bar p$) and $\gamma$--quanta.
The rather simple decay features allow an effective neutral or
charge-multiplicity--step trigger. It is understood, however, that the 
hyperons with quarks heavier than the strange quark do not have a unique decay 
scheme any more and the restriction to particular decay channels seems to be
appropriate.\\[-0.6cm]
\section{The System: 3 $\times s$  and 3 $\times \bar s$ quarks}
A systematic study of antihyperon-hyperon production with increasing strangeness
(charm) content is certainly interesting due to the change of the quark dynamics in
the different flavour environment of hadronic matter.\\
Besides the general interest of this research the production of the six--quark
system with three $s$--quarks and three $\bar s$--quarks in the final state (i.e.
the observation of the $\bar \Omega \Omega $ baryon pair and the 
$\phi \phi \phi$ three meson system)
should be stressed. The angular distribution of $\bar \Omega $ (or $\Omega $)
would be symmetric to 90$^0$ c.m. if the intermediate state is a compound like
system state of gluonic matter with definite quantum numbers. If, however, other
production mechanisms as boson exchange probabilities are relevant, the symmetry
via a pure intermediate gluonic matter state will be broken. It would be an
interesting and important result to see whether there is a correlation between the
initial $\bar p$ ($p$) and the final $\bar \Omega \Omega $.\\
\begin{table}
\begin{tabular}{crrrrrrrr}
\hline
    \tablehead{1}{r}{b}{Hyperon\\    }
  & \tablehead{1}{r}{b}{Quark-\\content}
  & \tablehead{1}{r}{b}{I}
  & \tablehead{1}{r}{b}{J$^{\pi}$}
  & \tablehead{1}{r}{b}{Mass ~ \\~~~~~~(MeV/c$^2$)}
  & \tablehead{1}{r}{b}{~~~Mean life ~ \\(s $\times 10^{-13}$)}
  & \tablehead{1}{r}{b}{$c\tau ~$ \\~~~~(cm)}   
  & \tablehead{1}{r}{b}{$\alpha_{Main}$ }   \\
\hline
$\Lambda~~~$ &uds &0             &$\frac{1}{2}$$^+$ &1115.57 $\pm$ 0.06   &2632 $\pm$ 20 &7.89  & +~0.642 $\pm$ 0.013   \\
$\Sigma^+~$   &uus &1             &$\frac{1}{2}$$^+$ &1189.37 $\pm$ 0.06   & 799 $\pm$ 4~~  &2.40  & -~0.980 $\pm$ 0.015   \\
$\Sigma^0~~$  &uds &1             &$\frac{1}{2}$$^+$ &1192.55 $\pm$ 0.10   &~~7.4 $\times 10~^{-7}$                        \\
$\Sigma^-~$   &dds &1             &$\frac{1}{2}$$^+$ &1197.50 $\pm$ 0.05   &1479 $\pm$ 11 &4.40  & -~0.068 $\pm$ 0.008   \\
$\Xi^0~~$     &uss &$\frac{1}{2}$ &$\frac{1}{2}$$^+$ &1314.80 $\pm$ 0.8~~  &2900 $\pm$ 90 &8.69  & -~0.411 $\pm$ 0.022   \\
$\Xi^-~~$     &dss &$\frac{1}{2}$ &$\frac{1}{2}$$^+$ &1321.34 $\pm$ 0.14   &1639 $\pm$ 15 &4.91  & -~0.456 $\pm$ 0.014   \\
$\Omega^-~~$  &sss &0             &$\frac{3}{2}$$^+$ &1672.43 $\pm$ 0.14   & 822 $\pm$ 12 &2.46  & -~0.026 $\pm$ 0.026   \\
\\
$\Lambda_c^+~~$    &udc &0              &$\frac{1}{2}$$^+$    &2285.2~~ $\pm$ 1.2~~ &1.91 $\pm$0.15 &0.006  &          \\
$\Sigma_c^{0}~~$   &ddc &1              &$\frac{1}{2}$$^+$    &2452.7~~ $\pm$ 1.3~~ &                &    &     \\
$\Sigma_c^{++}$    &uuc &1              &$\frac{1}{2}$$^+$    &2453.0~~ $\pm$ 1.2~~ &                &    &    \\
$\Sigma_c^{+}~$    &udc &1              &$\frac{1}{2}$$^+$    &2453.2~~ $\pm$ 1.2~~ &                &    &    \\
$\Xi_c^{+}~~$      &usc &$\frac{1}{2}$  &$\frac{1}{2}$$^+$(?) &2466.5~~ $\pm$ 2.5~~ &3.0 $\pm$ 1.0 &0.009  &         \\
$\Xi_c^{0}~~$      &dsc &$\frac{1}{2}$  &$\frac{1}{2}$$^+$(?) &2473.1~~ $\pm$ 2.0~~ &0.82 $\pm$ 0.6 &0.002  &      \\
$\Omega_c^{0}~~$   &ssc &$\frac{1}{2}$  &$\frac{1}{2}$$^+$(?) &2740.~~~~ $\pm$ 2.0~~ &       \\[0.1cm]
\hline
\\[-0.5cm]
\end{tabular}
\caption{Some properties of hyperons~[5]}
\label{tab:b}
\end{table}
\\[-0.5cm]
Since the asymmetry parameter $\alpha$ of the $\Omega$ decay is very small (see
table~2), its polarization
features cannot be measured via the weak decay directly. The determination of the
polarization of the weakly decaying daughter hyperons allows to extract spin
observables for the $\Omega$ particle. \\
A distinct feature of the $\bar \Omega \Omega$ production is the creation of two
spin 3/2 objects out of the $\bar p p$ interaction. Results from the PS185
experiment proove a clear dominance of the triplet $\bar s s $ production at
threshold. Since in the $\bar \Omega (\Omega)$ the three $s$--quarks ($\bar
s$--quarks) are oriented parallel, the three $\bar s s $ pairs created out of the
gluonic intermediate state should have spin quantum number as 3$^-$ if the 
$\bar \Omega (\Omega)$ is created with relative L~=~0 angular momentum.\\
The comparison of the $\bar \Omega (\Omega)$ baryon pair to the $\phi \phi \phi$ 
three meson production (where the three $I^G~(J^{PC})~=~0^-~(1^{--})$ $\bar s s$ mesons may not but can be
produced with no relative correlation) would give valuable information for a unique
determination of the intermediate matter state.\\[-0.2cm]

\end{document}